\documentclass[10pt,letterpaper]{article}
\usepackage{opex3,cite}
\usepackage{graphicx,color,textcomp}
\usepackage[english]{babel}
\usepackage{subfig}

\begin{document}

\title{Effects of inhomogeneities and drift on the dynamics of temporal solitons in fiber cavities and microresonators}

\author{P. Parra-Rivas$^{1,2}$, D. Gomila$^{2}$, M.A. Mat\'{\i}as$^{2}$, P. Colet$^{2}$, and L. Gelens$^{1,3}$}

\address{
$^1$Applied Physics Research Group (APHY), Vrije Universiteit Brussel (VUB),
Pleinlaan 2, 1050 Brussels, Belgium;\\
$^2$Instituto de F\'{\i}sica Interdisciplinar y
  Sistemas Complejos, IFISC (CSIC-UIB),Campus Universitat de les Illes
  Balears, E-07122 Palma de Mallorca, Spain;\\
$^3$Dept. of Chemical and Systems Biology, Stanford University School of 
Medicine, Stanford CA 94305-5174, USA.
}

\email{lendert.gelens@vub.ac.be} 



\begin{abstract}
In Ref.\ \cite{Parra_Rivas_1}, using the Swift-Hohenberg equation, we introduced a mechanism that allows to generate oscillatory and excitable soliton dynamics. This mechanism was based on a competition between a pinning force at inhomogeneities and a pulling force due to drift. Here, we study the effect of such inhomogeneities and drift on temporal solitons and Kerr frequency combs in fiber cavities and microresonators, described by the Lugiato-Lefever equation with periodic boundary conditions. We demonstrate that for low values of the frequency detuning the competition between inhomogeneities and drift leads to similar dynamics at the defect location, confirming the generality of the mechanism. The intrinsic periodic nature of ring cavities and microresonators introduces, however, some interesting differences in the final global states. For higher values of the detuning we observe that the dynamics is no longer described by the same mechanism and it is considerably more complex.
\end{abstract}



\section{Introduction}
Dissipative solitons \cite{Akhmediev} are (spatially and/or temporally) localized structures that have been shown to appear in a wide range of systems, ranging from chemical reactions \cite{Lee} to granular media \cite{Umbanhowar}, fluids \cite{Thual}, magnetic ferrofluids \cite{Barashenkov_PRL} and optics \cite{schapers_interaction_2000,barland_cavity_2002,leo_temporal_2010}. Dissipative solitons generated in nonlinear optical cavities are also called cavity solitons (CSs).  CSs in cavities with a cubic nonlinearity can be described by the well-known Lugiato-Lefever equation (LLE) \cite{lugiato_spatial_1987} and have been studied in great detail over the last decades. More recently, soliton dynamics in the LLE has received a renewed interest as they have been shown to play an important role in the stability and generation of optical Kerr frequency combs (KFCs) in microresonators \cite{Coen_1,Chembo_1}. Stable KFCs permit measuring light frequencies and time intervals with extraordinary accuracy, leading to numerous key applications \cite{kippen}. Various studies demonstrated that temporal CSs circulating within the optical microresonator correspond to KFCs at the output, and that their dynamical behavior immediately influences that of the KFCs \cite{leo-lendert,Parra_Rivas_2,Parra_Rivas_3, Chembo2}.

Effects of inhomogeneities and drift are present in many optical, chemical and fluid systems. In optical systems the drift can be produced by misalignments of mirrors \cite{santa,louver}, nonlinear crystal birefringence \cite{ward, santa2}, parameter gradients \cite{schapers} or by higher order effects chromatic light dispersion \cite{Parra_Rivas_3}. Inhomogeneities can originate from mirror or waveguide imperfections in an optical cavity and from the presence of fiber impurities, leading to variations in absorption coefficient or refractive index \cite{Kramper, Pedaci, Kozyreff}. Synchronously pumped fiber cavities have also been shown to be modeled by a LLE with a well-defined inhomogeneity in the pump \cite{Haelterman}.

Here, we focus on the dynamics of a single CS and its corresponding KFC in the presence of inhomogeneity and drift. The inhomogeneity will be introduced as a local change in the pump power in the LLE, while the drift will be modeled by a general gradient term. In Section 2, we introduce the LLE including terms accounting for inhomogeneity and drift. In Section 3, we discuss the bifurcation scenario in the presence of inhomogeneity and drift leading to oscillatory and excitable dynamics. We show that these dynamics are similar as in the Swift-Hohenberg equation (SHE) \cite{Parra_Rivas_1,Parra_Rivas_4}. Moreover we demonstrate how the dynamics of CSs that are periodically generated at the inhomogeneity are altered by the periodicity of the boundary conditions. Such boundary conditions allow those same CSs to interact with the defect again after having traveled one full roundtrip in the cavity. We also briefly show that the CS dynamics can be much more complex at higher values of the cavity detuning. Finally, in Section 4, we end with a short discussion.

\begin{figure}[tbp]
\centering
\includegraphics[width=8cm]{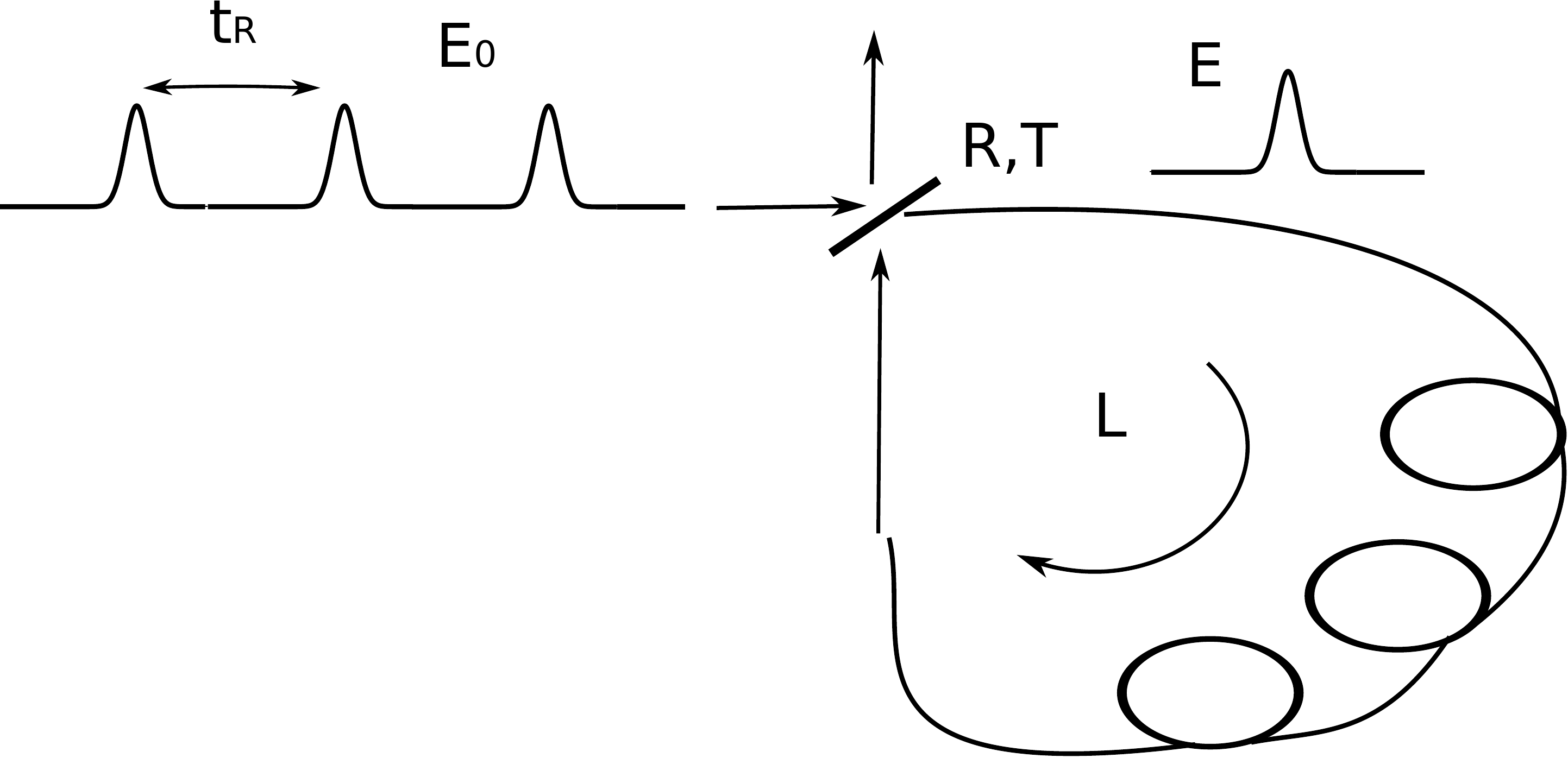}
\caption{A synchronously pumped fiber cavity. R and T are the reflection and transmission coefficients of the beam splitter. L is the length of the fiber.}
\label{sketch}
\end{figure}

\section{The Lugiato-Lefever equation with inhomogeneity and drift}
{We consider a prototypical setup that can illustrate the various dynamical effects triggered by the competition between inhomogeneities and drift.} Fig.\ref{sketch} shows a fiber cavity of length $L$, a beam splitter with reflection and transmission coefficients $R$ and $T$, and a source such as a mode-locked laser that emits a train of pulses of amplitude $E_0$.  At the beam splitter, this pulse train is added to the electromagnetic wave circulating inside the fiber at the beam splitter. Here we consider that the cavity is synchronously pumped by a periodic train of pulses $E_0(\tau)$ at a repetition frequency $1/t_R$, where $t_R$ is the round-trip time of the cavity given by $t_R=L/c$ (with $c$ the speed of light in the medium). We also assume that the pulse duration is much shorter than the round-trip time $t_R$. The evolution of the optical field $E=E(t,\tau)$ within the cavity after each round-trip is described by the following equation \cite{Haelterman}:
\begin{equation}
 t_R\frac{\partial E}{\partial t}=-(\alpha+i\delta_0)E-i\frac{L\beta_2}{2}\frac{\partial^2E}{\partial\tau^2}
 +i\gamma L |E|^2E +\sqrt{T}E_0(\tau),
\label{eq3}
\end{equation}
where $t_R$ is the round-trip time, $\alpha$ describes the total cavity losses, $\beta_2$ is the second order dispersion coefficient, $\gamma$ is the nonlinear coefficient due to the Kerr effect in the resonators, and $\delta_0$ is the cavity detuning. The slow time $t$ describes the wave evolution after each round-trip and $\tau$ is the
fast time describing the temporal structure of the nonlinear waves. Due to the periodicity of the driving, the system is described with periodic boundary conditions. After normalizing Eq.\ \ref{eq3} we arrive to the dimensionless mean-field LLE \cite{lugiato_spatial_1987}:
\begin{equation}\label{eq.4}
 \displaystyle\frac{\partial u(t',\tau')}{\partial t'}=-(1+i\theta)u(t',\tau')+i|u(t',\tau')|^{2}u(t',\tau')+i\displaystyle\frac{\partial^2 u(t',\tau')}{\partial\tau'^2}+u_{0}(\tau'),
\end{equation}
with $t'=\alpha t/t_R$, $\tau'=\tau\sqrt{2\alpha/(L|\beta_2|)}$, $u(t',\tau')=E(t,\tau)\sqrt{\gamma L/\alpha}$, $u_0(\tau)=E_0(\tau)\sqrt{\gamma L\theta/\alpha^3}$,
and $\theta=\delta/\alpha$. We will not consider any potential higher order dispersion effects as analyzed in Refs.\ \cite{Gelens_PRA_2007,Gelens_PRA_2008,GelensOL2010}. In what follows we will drop the primes in the notation of $t'$ and $\tau'$ and we choose a normalized $\tau-$domain of length $L_{\tau} = 70$. In many cases the pump can be approximated by a continuous wave (cw) $u_0(\tau)=u_0$. However, when the amplitude of the pulse train is not negligible, this cw approximation is no longer valid and every round-trip the pump changes in intensity. We introduce such inhomogeneity in the pump by approximating each pulse by a Gaussian function of amplitude $h$ and width $\sigma$ centered at $\tau = \tau_0$:
\begin{equation}\label{eq.defect}
u_0(\tau) = u_0+b(\tau) = u_0 + h \exp\left(-\left(\frac{\tau-\tau_{0}}{\sigma}\right)^{2}\right).
\end{equation}

There exist various sources of drift in fiber cavities and microresonators. Any odd order of chromatic dispersion breaks the reversibility ($\tau \rightarrow -\tau$) in the LLE and induces a drift of CSs. For example, third order dispersion effects ($\sim d_3\partial^3_\tau$) can play an important role in dispersion compensated cavities ($\beta_2\approx0$), where it has been shown to stabilize KFCs \cite{Parra_Rivas_2}. In order to cover, at least qualitatively, the effect of all possible sources of drift, we model such drift by adding a general gradient term $-c\partial u / \partial \tau$ to Eq.\ (\ref{eq.4}):
\begin{equation}\label{eq.5}
 \displaystyle\frac{\partial u}{\partial t}=-(1+i\theta)u+i|u|^{2}u-c \displaystyle\frac{\partial u}{\partial\tau}+i\displaystyle\frac{\partial^2 u}{\partial\tau^2}+u_{0}(\tau) ,
\end{equation}

In the following Section, we will explore the dynamics of a single CS in the LLE (\ref{eq.5}) for small values of the frequency detuning ($41/30 <  \theta < 2$). In this region, in the absence of inhomogeneities and drift, a pattern is created at a Modulational Instability (MI) at $I_s =  |u_{s}|^{2} = 1$, where $u_s$ is the homogeneous steady state (HSS). {Below threshold ($I_s < 1$), where CSs exist, their profile away from the core approaches the HSS in an oscillatory way. For larger values of the detuning, the MI no longer exists and, increasing the pump, the wavelength of the CS tails oscillations quickly increases, leading to very smooth or no oscillatory tails. CSs have been shown to be always stable in the low frequency detuning region in the LLE without drift and inhomogeneities \cite{gomila-scroggie}, which is why we focus on analyzing the effect of drift and defect in this region of operation. For higher values of the frequency detuning various dynamical regimes such as oscillations and chaos have been reported \cite{Barashenkov_PRE, leo-lendert, Parra_Rivas_2}.  A more detailed study of how these rich dynamics are influenced by drift and inhomogeneities is left for future work. 

\section{Dynamical regimes}
In this Section, unless mentioned otherwise, we fix the values $\theta=1.56$ and $u_0=1.137$ within the 
low frequency detuning region ($41/30 < \theta < 2$), and such that single CSs 
exist in the LLE without drift and inhomogeneity. We also choose 
$\sigma=0.2727$ around half the width of the CS at half maximum and 
$\tau_0=t_R/2$, such that the inhomogeneity is centered in the $\tau-$domain. 
Similar behavior can be found for other values of $\theta$ and $u_0$ within this 
region. 

\begin{figure}
\centering
\includegraphics[scale=1]{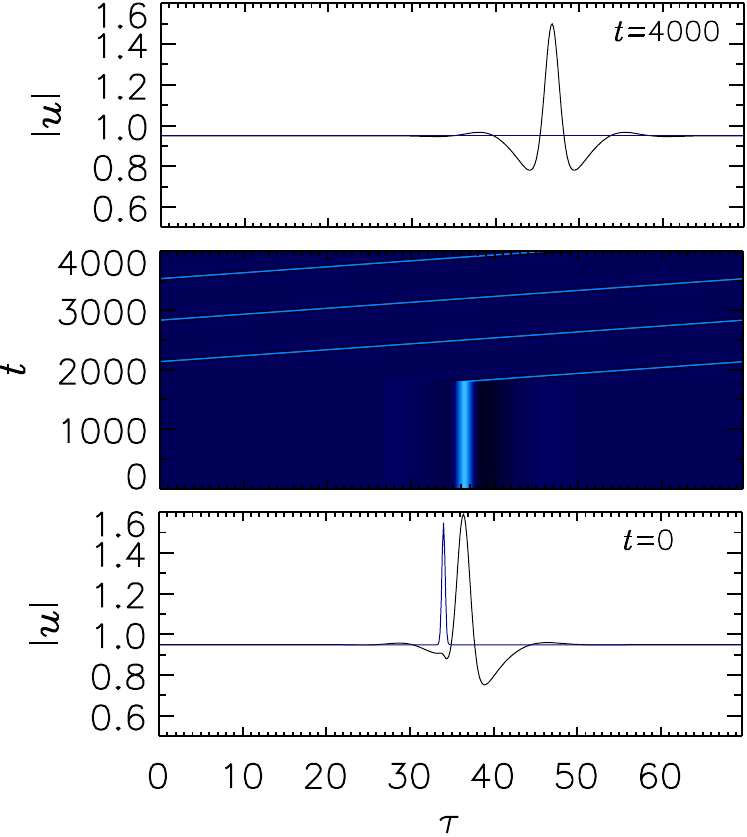}
\caption{Effect of a defect on drifting CS solutions. Parameters for $t = 0 - 1800$ are $h=0.6$, $c=0.1$.  At $t=1800$, $h$ is set to $0$ and then the soliton starts drifting. The time evolution is shown in the middle panel, while the initial and final CS profiles are plotted in the bottom and top panels in black. The corresponding profile of the inhomogeneity is added in blue. Other parameters are $\theta=1.56$, $u_0=1.137$, $\sigma=0.2727$ and $\tau_0=t_R/2$.}
\label{c0}
\end{figure}

Fig.\  \ref{c0} illustrates the competition between the pulling force of the drift term (c=0.1) and the pinning force of the defect (h=0.6). {Initially the pinning force of the defect is strong enough to prevent the CS from moving and it remains anchored close to the defect.} When the defect is removed the solution starts to drift with its drift speed determined by the strength $c$. The CS profile is not altered by the drift term ($c \neq 0, h=0)$ and merely moves with fixed speed. In the presence of a defect ($c \neq 0, h \neq 0$), it is clear that the defect deforms the whole CS profile, and this more strongly at locations where the amplitude of this inhomogeneity is highest. 

\begin{figure}       
\centering                
\includegraphics[scale=1]{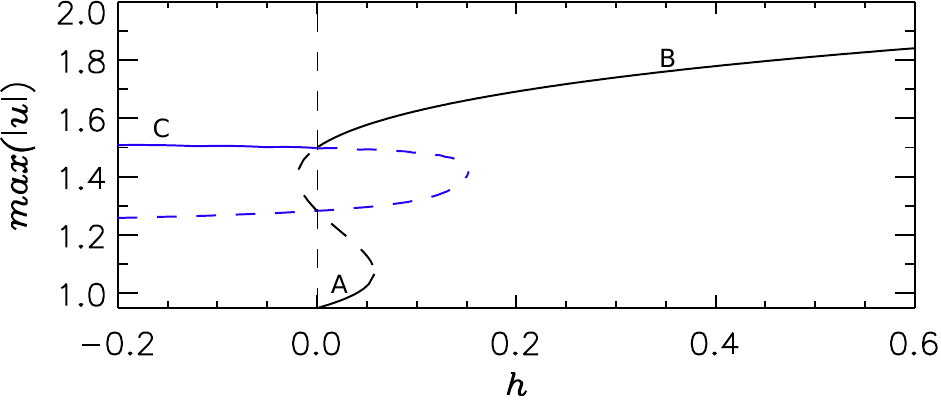}
\caption{Bifurcation diagrams of the different steady-state state solutions in function of $h$ with $c = 0$. The solid (dashed) lines represent the energy of the stable (unstable) states. Other parameters are as in Fig.\  \ref{c0}.}
\label{c0.0_LLE}
\end{figure}

\begin{figure}       
\centering                
\includegraphics[width=12cm]{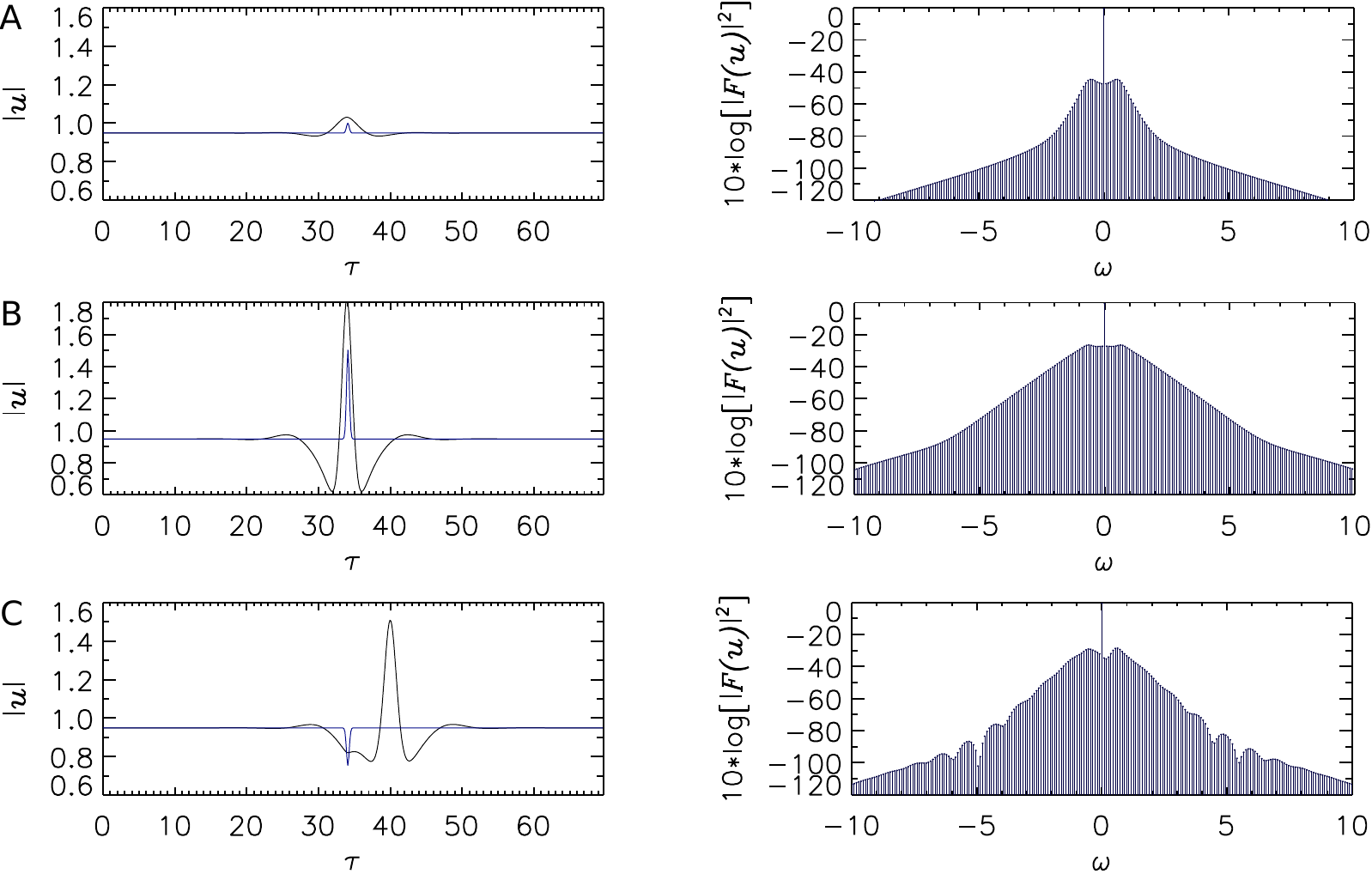}
\caption{Profiles of the different steady-state state solutions corresponding to the branches shown in Fig.\  \ref{c0.0_LLE}. The panels on the left depict the absolute value of the field $u$ inside the cavity (black) and the corresponding profile of the defect (blue), while the panels on the right show the corresponding KFC in dB scale. A. h =0.052, B. h = 0.556, and C. h = - 0.196. Other parameters are as in Fig.\  \ref{c0} and $c=0$.}
\label{states_sin_drift}
\end{figure}
 
Fig.\  \ref{c0.0_LLE} shows the bifurcation diagram of the steady-state solutions in the presence of a defect ($h \neq 0$), but without a drift term ($c = 0$) in more detail. We plot the maximum absolute value of the field $u$ as a function of $h$. Depending on the amplitude $h$ of the inhomogeneity, several pinned steady states appear. The fundamental state of the system (branch A) is now a small bump solution induced by the inhomogeneity rather than a perfect homogeneous solution (see Fig.\  \ref{states_sin_drift}A). When increasing the value of $h$ the system reaches a high amplitude CS (branch B) pinned at its center (see Fig.\  \ref{states_sin_drift}B). Finally, {for negative values of $h$, CSs in branch C are pinned at the first oscillation of its tail (see Fig.\  \ref{states_sin_drift}C)}. {In the latter case a CS can pin at either side of the defect.}

\begin{figure}
\centering
\includegraphics[scale=1]{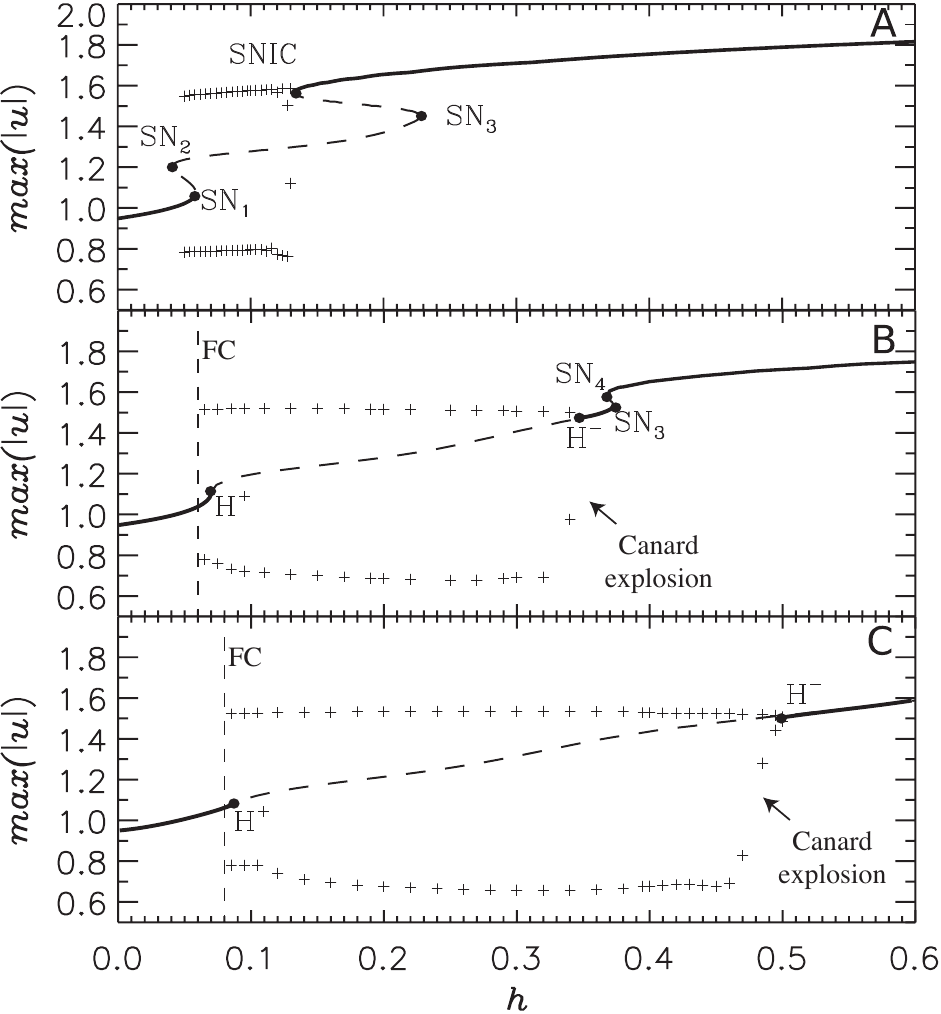}
\caption{Bifurcation diagrams of the different steady-state state solutions in function of $h$ for different values of the the drift strength $c$: A. $c=0.025$, B. $c=0.06$ and C. $c=0.1$. The solid (dashed) lines represent the energy of the stable (unstable) states. The $+$ markers correspond to the extrema of oscillatory solutions, and the vertical dashed line shows the location of the Fold of Cycles (FC). Other parameters are as in Fig.\  \ref{c0}.}
\label{dia2}
\end{figure}

When the drift term is taken into account ($c \neq 0$), the pinned states shown in Fig. \ref{c0.0_LLE} experience a force trying to detach them from the inhomogeneity. This competition between the inhomogeneity that pins the states to a fixed position and the drift force trying to pull them out, leads to the appearance of rich variety of dynamics as shown in Ref.\ \cite{Parra_Rivas_1}. Those dynamics comprised small and large amplitude oscillations (train of solitons) and soliton excitability. In Fig.\ref{dia2} we show how the bifurcation scenario in Fig.\ref{c0.0_LLE} changes with increasing values of the drift strength $c$. 

For low values of $c$ two extra saddle-node bifurcations appear involving 
unstable steady state solutions (see Fig.\ref{dia2}A). Moreover, a limit cycle 
is created in a SNIC (saddle node on the invariant circle) bifurcation. This 
limit cycle corresponds to a periodic generation and emission of CSs from the 
inhomogeneity resulting in a sequence of drifting solitons called train of 
solitons or soliton tap \cite{Caboche}. An example of such a train of solitons 
(for a higher value of $c$) is shown in Fig.\ref{fig:dynamics_c0.1}A. At the 
SNIC the period of emission of CSs diverges, and it decreases as one moves away 
from the SNIC bifurcation point \cite{jacob}. In the study of the SHE in Ref.\ 
\cite{Parra_Rivas_1}, such oscillations were also observed, but the boundary 
conditions were chosen to be absorbing such that the emitted solitons 
disappeared at those boundaries. Here, due to the periodic boundary conditions, 
the train of solitons are instead reinjected on the other side of the domain, 
filling up the whole cavity. 

\begin{figure}
  \centering
\includegraphics[width=12cm]{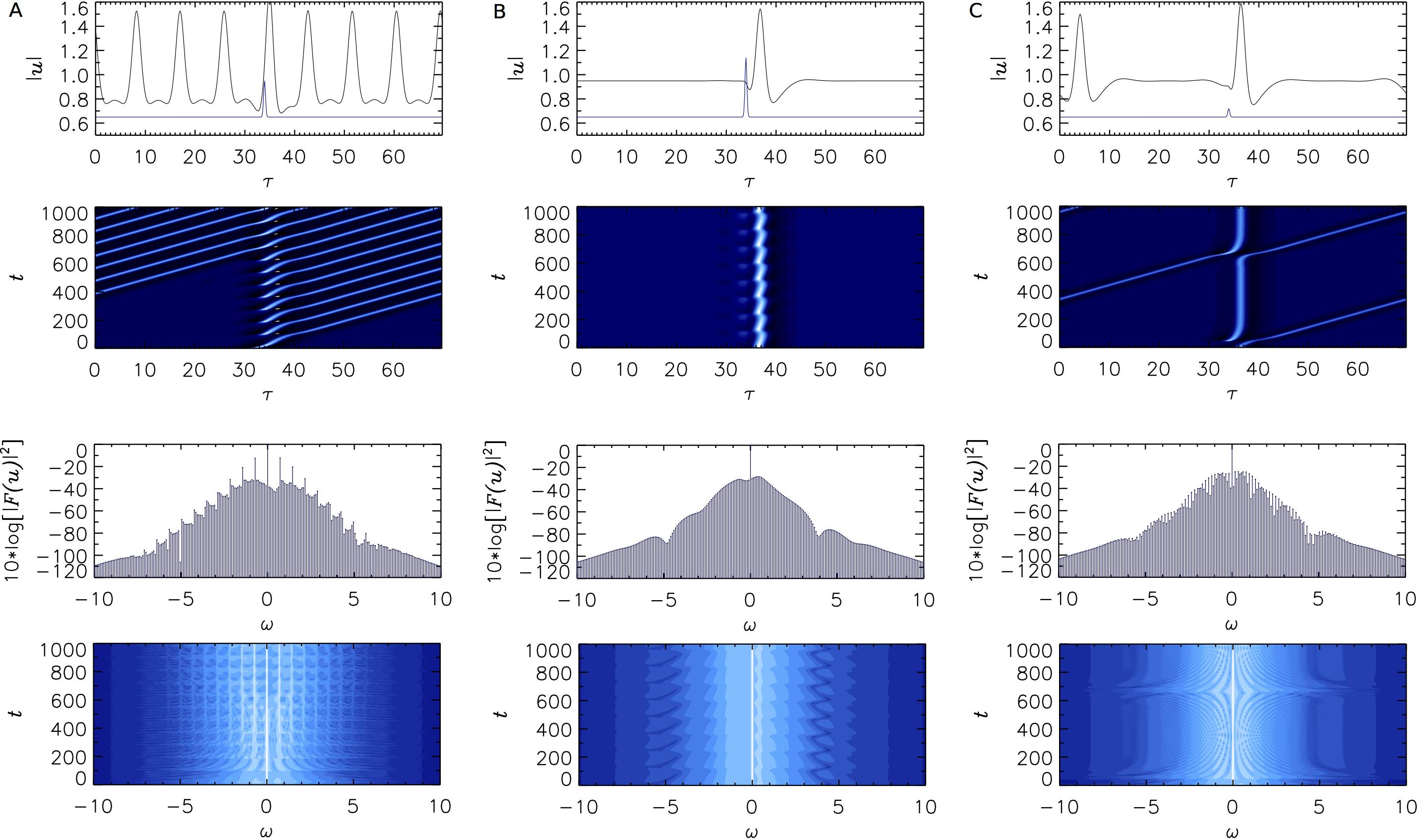}
\caption{Evolution of a CS at A. $c=0.1$, $h=0.3$ (large-amplitude oscillations), B. $c=0.1$, $h=0.49$, (small-amplitude oscillations), and C. $c=0.1$, $h=0.3$ (excitability). The top two panels depict the evolution and final profile of the absolute value of the field $u$ inside the cavity, while the bottom two panels show the corresponding evolution and final profile of the corresponding KFC in dB scale. Other parameters are as in Fig.\  \ref{c0}. In panel A, a periodic train of solitons is created at the inhomogeneity. In panel B, a soliton is pinned at the inhomogeneity and locally oscillates with small amplitude. In panel C, the system is excited by transiently ($\Delta t=30$) changing the parameter values by $h \rightarrow h + \Delta h$, with $\Delta h = - 0.3$. Such a parametric excitation leads to the emission of a CS form the defect location that continues to circulate in the cavity.}
\label{fig:dynamics_c0.1}
\end{figure}

For higher values of $c$, two saddle-node bifurcations $SN_1$ and $SN_2$ collide 
in a cusp bifurcation, a subcritical Hopf bifurcation $H^+$ appears, the SNIC 
disappears in favor of another saddle-node bifurcation $SN_4$, and a 
supercritical Hopf bifurcation $H^-$ is created (see Fig.\ref{dia2}B). Finally, 
increasing the value of $c$ further (see Fig.\ref{dia2}C), all saddle-node 
bifurcations have disappeared and a single branch remains with a supercritical 
$H^-$ and subcritical $H^+$ Hopf bifurcation. In $H^-$, a stable limit cycle is 
created. This limit cycle initially corresponds to oscillations of small 
amplitude which remain localized at the defect position in the cavity (see 
Fig.\ref{fig:dynamics_c0.1}B). When decreasing the strength of the 
inhomogeneity, these oscillations rapidly increase in amplitude in a so-called 
Canard explosion \cite{canard} and lead to the detachment of solitons from the 
defect. Those solitons then drift away and lead to a train of solitons that 
continue to circulate inside the cavity (see Fig.\ref{fig:dynamics_c0.1}A). 
These large-amplitude oscillations persist until a Fold of Cycles (FC) 
bifurcation where the stable limit cycle collides with an unstable limit cycle 
originating at $H^+$. At values of the defect strength $h$ beyond the 
supercritical $H^-$ bifurcation, the pinned CS is stable. However, the system 
can be excited to emit a CS from the defect location through direct perturbation 
of the CS profile or by transiently changing the parameter to the nearby 
oscillatory regime as shown in Fig.\  \ref{fig:dynamics_c0.1}C.

{The unfolding of these various bifurcations is formally identical to the ones found in the context of the SHE. The physical mechanism behind this scenario is the interplay between the pinning of a CS to an inhomogeneity and the drift. The LLE admits stable localized solutions that have characteristic oscillatory tails similar to those of the SHE. This thus seems to be a sufficient condition to display the same CS dynamics.  
For a more detailed bifurcation analysis we refer to the work on the SHE, presented in Refs.\ \cite{Parra_Rivas_1,Parra_Rivas_4}. In these references, drifting CSs emitted from the defect were absorbed at the boundary of the domain. Here, periodic boundary conditions lead to the recirculation of CSs in the domain and to interaction with the defect at the center of the domain and with the new CS emitted there. 
In particular, Fig.\  \ref{period_change} demonstrates that the large-amplitude oscillations ($h=0.3$ and $c=0.1$) are modified through this interaction. The temporal evolution of the field $u$ at the defect position is plotted for both absorbing (black line) and periodic (red line) boundary conditions. In the case of absorbing boundary conditions, the period of the oscillations $T_0$ is essentially given by the Hopf frequency. The spatial wavelength of the emitted train of solitons is therefore $\lambda_0 \simeq c T_0$. In Fig.\  \ref{period_change}A one can see that the shape of the oscillations is slightly adjusted as soon as the first emitted CS reaches the defect after one roundtrip. In this case the period of the oscillations does not change considerably because the train of CSs (consisting of $n$ peaks) emitted by the defect with a natural period $T_0$ has a wavelength $\lambda_0$ that is an almost exact submultiple of the cavity length $L_{\tau} \approx n \lambda_0$. In general this will not be the case for arbitrary cavities. Fig.\ \ref{period_change}B shows the comparison of the temporal evolution with absorbing and periodic boundary conditions for a domain size of $L_{\tau} \approx 85$. The same amount of peaks ($n=9$) are emitted but now since a multiple of the natural wavelength no longer fits exactly within the cavity length, the period of the oscillations changes more considerably.}

\begin{figure}
\centering
\includegraphics[width=12cm]{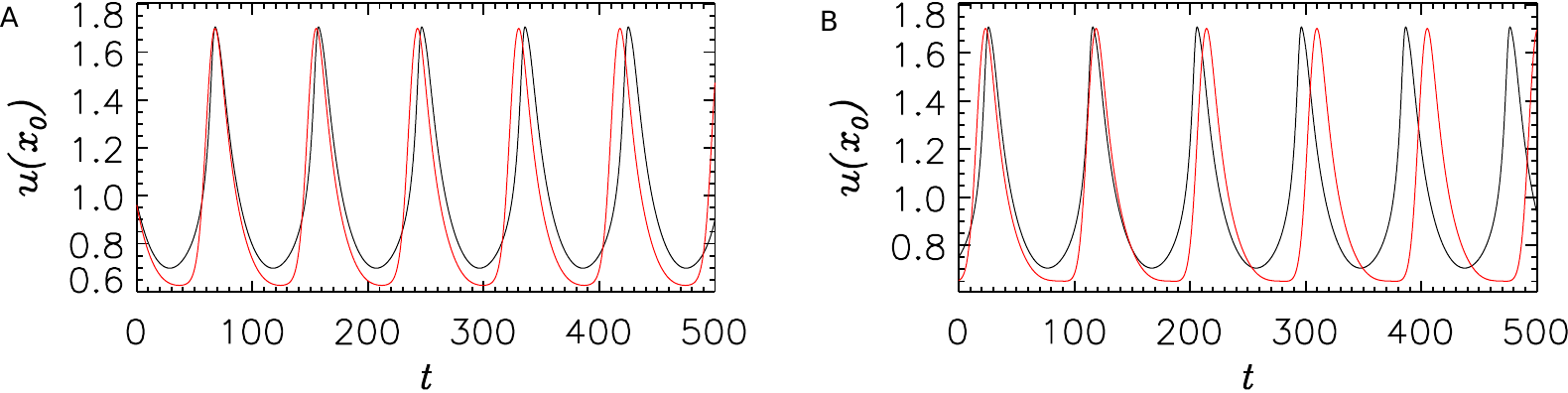}
\caption{Temporal evolution of the field $u$ at the defect position $\tau_0$ for two different boundary conditions: 1) absorbing boundary conditions (black solid line), 2) periodic boundary conditions (red dashed line). The train of solitons generated in the cavity corresponds to 9 peaks that circulate (branch H in Fig.\ \ref{curva}). In panel A the domain width $L_{\tau} = 78$, while in panel B $L_{\tau} = 84.8$. $h=0.3$ and $c=0.1$, while other parameters are as in Fig.\  \ref{c0}. }
\label{period_change}
\end{figure}

\begin{figure}
\centering
\includegraphics[scale=1]{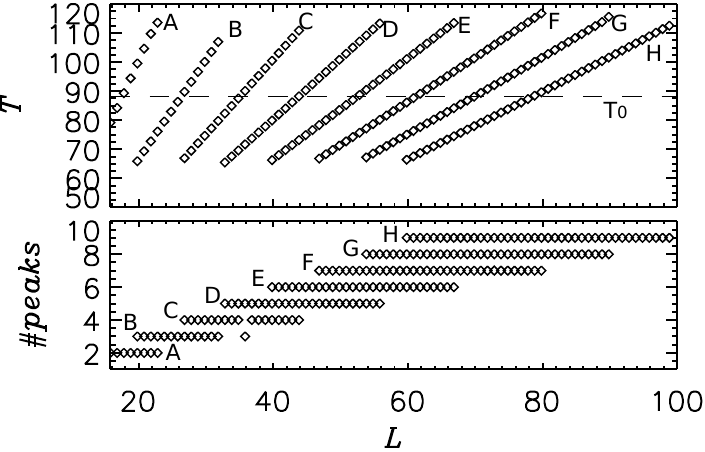}
\caption{The top panel shows the oscillation period $T$ of the various solutions of train of solitons (A - H) in the system with periodic boundary conditions. The natural period $T_0$ in the system with absorbing boundary conditions is plotted in dashed lines as reference. The bottom panel shows the amount of CSs within the train of solitons corresponding to the branches shown in the top panel.}
\label{curva}
\end{figure}

Fig.\ \ref{curva} shows a more detailed analysis of how the period of oscillations changes when varying the cavity length $L_{\tau}$ in the presence of periodic boundary conditions. The top panel shows the oscillation period $T$ of the various solutions of train of solitons (A - H), while the natural period $T_0$ is plotted in dashed lines as reference. The bottom panel shows the amount of CSs within the train of solitons corresponding to the branches shown in the top panel. Each solution branch (A-H) increases its period with increasing domain sizes as the system tries to accommodate the same amount of peaks in the increasingly large domain. As the cavity becomes larger more solution branches coexist and the system allows for soliton trains with different amounts of CSs.

{The periodic boundary conditions also have an important effect in the excitable regime encountered beyond $H^-$, shown in Fig. 6C.}  The excitation 
leads to one CS that remains pinned in the defect and another CS that drifts 
away from the defect. Due to the periodic boundary conditions of the system, 
this drifting CS eventually collides with the pinned CS from behind. This 
collision frees the pinned CS from the defect such that it drifts away, while 
the CS that was previously drifting now takes its place and remains pinned at 
the location of the defect. This type of dynamics reminds of the classic 
Newton's cradle.

\begin{figure}       
\centering                
\includegraphics[width=12cm]{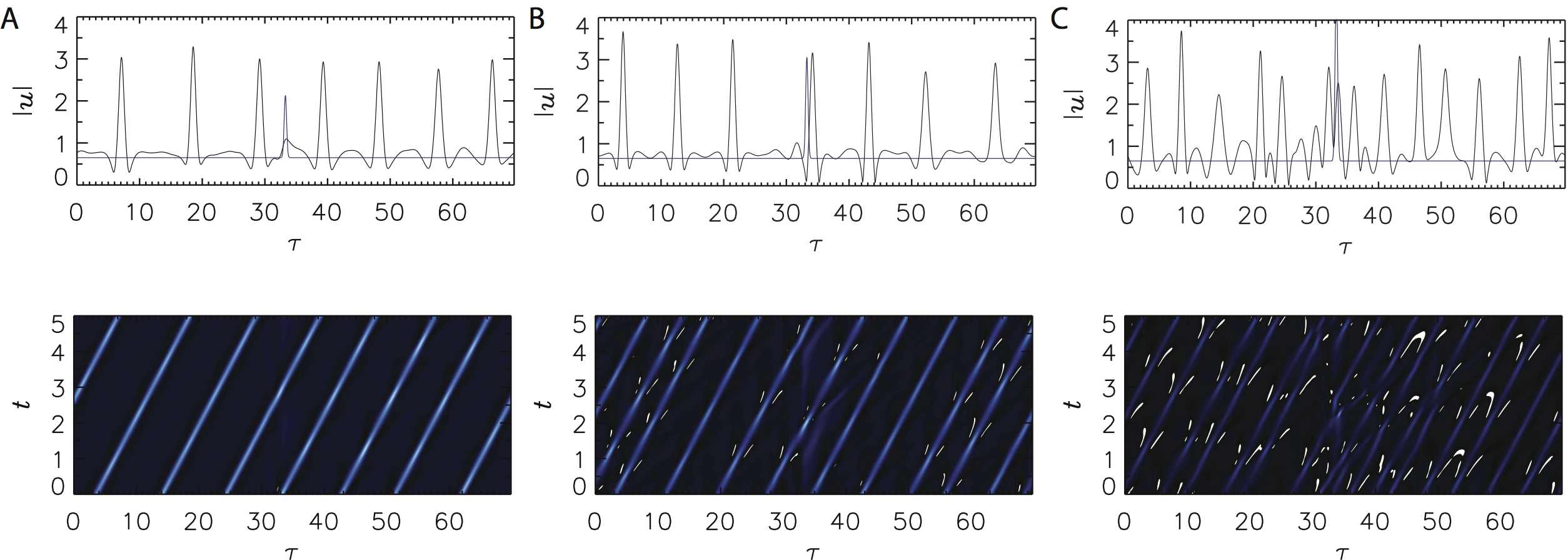}
\caption{Evolution of a train of solitons for A. $h = 1.482$, B. $h = 2.682$ and C. $h = 5.682$. The panels depict the evolution (bottom) and final profile (top) of the absolute value of the field $u$ inside the cavity. Other parameters are $\theta=3.8$, $u_0=2.6$, $c = 3$, $\sigma=0.2727$ and $\tau_0=t_R/2$.}
\label{dynamics_highdet}
\end{figure}

Finally, Fig.\ \ref{dynamics_highdet} illustrates how the dynamics can be altered for higher values of the frequency detuning. Here, we choose ($\theta=3.8$, $u_0=2.6$), values close to a Hopf instability in the LLE without drift and defect. At such higher values of $\theta$ the LLE without defect and drift has been shown to exhibit a wide range of oscillatory dynamics \cite{Barashenkov_PRE, leo-lendert, Parra_Rivas_2}. Trains of solitons still exist and can originate from an oscillatory instability at the defect location that emits one CS at a time. Similar as in Fig.\ \ref{fig:dynamics_c0.1}A, the CSs eventually fill up the whole domain and continue to circulate in the cavity. However, in contrast to the stable drifting trains of solitons in the low detuning region (Fig.\ \ref{fig:dynamics_c0.1}A), these solutions can now undergo a wide range of instabilities. In Fig.\ \ref{dynamics_highdet}A each CS within the train of solitons oscillates, but not necessarily with the same frequency. For higher values of the defect strength $h$, the trains of solitons start behaving more chaotically, see Fig.\ \ref{dynamics_highdet}B-C. A detailed analysis of the origin and organization of these various instabilities at higher values of $\theta$ is beyond the scope of this work and will be investigated in the future.

\section{Discussion}
In this work we extended the analysis presented in Ref.\ \cite{Parra_Rivas_1}, 
where it was shown that the competition between inhomogeneities and drift can 
lead to oscillations and excitability of temporal solitons in the  
Swift-Hohenberg equation. In the context of nonlinear optical cavities such as 
fiber cavities and microresonators, we have shown that a similar bifurcation 
scenario leads to the periodic emission of cavity solitons from locations in the 
cavity containing defects or imperfections. This confirms the generic nature of 
this dynamics and argues that the main ingredients for the generation of trains 
of solitons are a) inhomogeneities that can exert a pinning force on the 
soliton, b) a drift that gives rise to a pulling force on the soliton, and c) 
soliton solutions with oscillatory tails. We have also shown that as the cavity 
detuning $\theta$ is increased in the LLE, those oscillatory tails become much 
weaker and more complex dynamical behavior appears. In future work, the 
effect of multiple and/or unstructured defects, the dynamics of bound states of 
CSs, and the bifurcation scenario at higher values of the cavity detuning can be studied. 

Inhomogeneities and drift in fiber cavities and microresonators are unavoidable due to imperfections in the fabrication process, material properties and higher order chromatic light dispersion. Therefore, we believe that the type of dynamics studied in this work could be of considerable importance for all applications based on temporal solitons in nonlinear optical cavities. Optical Kerr frequency combs have especially proven to be important for a wide range of applications \cite{kippen}. Since temporal cavity solitons have been shown to lie at the heart of such Kerr frequency combs, instabilities of such solitons induced by imperfections of the microresonator can be particularly relevant for all applications relying on stable frequency combs.

\section{Acknowledgments}
This research was supported by the Research Foundation - Flanders (L.G. and P. 
P.-R.), by the research council of the VUB, by the Belgian American Educational 
Foundation (L.G.), by the Belgian Science Policy Office (BelSPO) under Grant 
No. IAP 7-35, and by the Spanish MINECO and FEDER under Grant INTENSE@COSYP 
(FIS2012-30634), and the Comunitat Aut\`onoma de les Illes Balears.
\end{document}